\documentclass[pmlr]{jmlr_arxiv}

\usepackage{longtable}

\usepackage{booktabs}
\usepackage[load-configurations=version-1]{siunitx} 

\jmlrvolume{}
\jmlryear{}
\jmlrworkshop{}

\usepackage[utf8]{inputenc} 
\usepackage[T1]{fontenc}    
\usepackage{hyperref}       
\hypersetup{
  colorlinks   = true, 
  urlcolor     = blue, 
  linkcolor    = black, 
  citecolor   = black 
}

\usepackage{acro}
\DeclareAcronym{AI}{short = AI, long = artificial intelligence}
\DeclareAcronym{AMM}{short = AMM, long = automated music mastering}
\DeclareAcronym{CNN}{short = CNN, long = convolutional neural network}
\DeclareAcronym{DAW}{short = DAW, long = digital audio workstation}
\DeclareAcronym{MIR}{short = MIR, long = music information retrieval}
\DeclareAcronym{MSS}{short = MSS, long = music source separation}
\DeclareAcronym{NLP}{short = NLP, long = natural language processing}
\DeclareAcronym{LLM}{short = LLM, long = large language model}
\DeclareAcronym{XAI}{short = XAI, long = explainable artificial intelligence}
\DeclareAcronym{VST}{short = VST, long = virtual studio technology}

\title[Prevailing Research Areas for Music AI in the Era of Foundation Models]{Prevailing Research Areas for Music AI \\ in the Era of Foundation Models}

 \author{\Name{Megan Wei} \Email{meganwei@brown.edu}\\
 \addr Brown University
 \AND
 \Name{Mateusz Modrzejewski} \Email{mateusz.modrzejewski@pw.edu.pl}\\
 \addr Warsaw University of Technology
 \AND  
 \Name{Aswin Sivaraman} \Email{asivara@iu.edu}\\
 \addr Indiana University 
 \AND
 \Name{Dorien Herremans} \Email{dorien\_herremans@sutd.edu.sg}\\
 \addr Singapore University of Technology and Design 
}

\begin{document}

\maketitle

\begin{abstract}
Parallel to rapid advancements in foundation model research, the past few years have witnessed a surge in music AI applications. As AI-generated and AI-augmented music become increasingly mainstream, many researchers in the music AI community may wonder: what research frontiers remain unexplored?

This paper outlines several key areas within music AI research that present significant opportunities for further investigation. We begin by examining foundational representation models and highlight emerging efforts toward explainability and interpretability. We then discuss the evolution toward multimodal systems, provide an overview of the current landscape of music datasets and their limitations, and address the growing importance of model efficiency in both training and deployment.

Next, we explore applied directions, focusing first on generative models. We review recent systems, their computational constraints, and persistent challenges related to evaluation and controllability. We then examine extensions of these generative approaches to multimodal settings and their integration into artists' workflows, including applications in music editing, captioning, production, transcription, source separation, performance, discovery, and education.

Finally, we explore copyright implications of generative music and propose strategies to safeguard artist rights. While not exhaustive, this survey aims to illuminate promising research directions enabled by recent developments in music foundation models.
\end{abstract}



\section{Introduction}\label{sec:introduction}

\Ac{MIR} is a mature interdisciplinary research area that has seen significant progress across many fronts, from model architectures and advanced music tagging to broadening coverage of global musical cultures.
However, the emergence of foundation models has transformed the landscape, capturing the attention of researchers, artists, and the public alike.
Major news outlets are now covering AI-powered tools for music composition as well as AI-generated music (e.g., the ``audio deepfakes'' of high-profile artists) \citep{ChowAR2023time}.
Moreover, music AI startups such as Suno, Udio, and Producer have rapidly emerged, offering prompt-based music synthesis, acoustic style transfer, and symbolic music control. These tools have accelerated creative workflows, enabling artists to iteratively explore new sounds and melodies as part of their ideation process.
Rapid improvements in foundation models have achieved capabilities previously considered unattainable. This immense progress might suggest that most challenges in MIR and music AI have been solved. Yet these same advances introduce novel paradigms that reveal new research directions and unresolved questions, fundamentally reshaping what problems we should address.

In this work, we explore emerging research directions in MIR, emphasizing the need for rigorous validation and integrity. While space constraints prevent comprehensive coverage, we highlight critical open problems and opportunities across these three categories: fundamental, applied, and responsible music AI. Our goal is to help the community rethink what is possible in this era and inspire potential answers to the age-old researcher's dilemma: \textbf{\textit{what should I work on now?}}

\section{Fundamental Music AI}\label{sec:fundamental}

\subsection{Music encoders}\label{sec:architecture}

In computer vision, ResNets \citep{he2016deep} and vision transformer models \citep{dosovitskiy2020image} have learned rich visual representations from extensive image datasets.
In natural language processing, BERT \citep{devlin2018bert} demonstrated large-scale language understanding, while the T5 transformer \citep{raffel2020exploring} introduced a unified text-to-text interface, paving the way for modern \acp{LLM}.

However, research on music foundation models has only recently gained momentum \citep{ma2024foundation}.
A music foundation model would need to understand music in multiple dimensions -- melody, harmony, rhythm, timbre, and diverse stylistic and cultural traits -- yielding universal representations for many downstream tasks. The Holistic Evaluation of Audio Representations (HEAR) challenge at NeurIPS 2021 \citep{turian2022hear} was among the first large-scale initiatives towards robust audio representations, providing a benchmark of 19 diverse tasks. 

Generatively pre-trained latent representations from models like Jukebox \citep{dhariwal2020jukebox} and MERT \citep{Li2023-dg} have proven useful for a variety of tasks, as demonstrated by evaluations on the MARBLE Benchmark \citep{yuan2023marble}.
Neural audio codecs such as EnCodec \citep{defossez2022highfi}, Descript \citep{kumar2023high} and SoundStream \citep{zeghidour2021soundstream} employ discrete neural representation learning \citep{van2017neural} to encode audio into discrete tokens, useful for generative tasks, like MusicGen \citep{copet2024simple} and MusicLM \citep{agostinelli2023musiclm}.
Finally, \cite{won2024foundation} and \cite{gardner2023llark} augment foundation models with music understanding.

Developing a unified music generation and understanding pipeline as well as building more efficient and robust architectures remains an open and activate research question.


\subsection{Explainability}\label{sec:xai}

Explainable artificial intelligence (XAI), essential for developing trustworthy and interpretable systems, has seen limited adoption in \ac{MIR}. Early work by \cite{ChoiK2016explaining, WonM2019towards} visualized and auralized spectrogram features to explain music tagging models, analogous to saliency maps in image processing. \cite{FoscarinF2022ismir} introduced a concept-based approach for explaining composer classification, that prioritized interpretability for musicologists rather than solely AI experts. \cite{MishraS2017ismir} adapted the LIME framework \citep{RibeiroMT2016kdd} to music content analysis and later to singing voice detection \citep{MishraS2020ijcnn}, while \cite{HaunschmidV2020towards} proposed audioLIME, which replaces image segmentation with source separation. Despite these efforts, adaptation of broader \ac{XAI} tools such as Shapley additive explanations (SHAP) \citep{LundbergSM2017nips}, layerwise relevance propagation (LRP) \citep{BachS2015plos}, and concept relevance propagation (CRP) \citep{AchtibatR2023nature} have seen sparse adoption, particularly beyond music tagging.

A newer direction, driven by generative models focuses on explanation through data attribution, identifying which musical content shapes system outputs. \cite{barnett2024exploring} study embedding similarities between prompts and outputs in VampNet \citep{garcia2023vampnet}, while \cite{batlle2024towards} propose metrics and tools to detect data replication.

\subsection{Interpretability}\label{sec:interp}


Several studies in language modeling have examined whether model representations encode meaningful features useful for downstream tasks \citep{Meng2022-id,tenney2019learn}, spanning domains from color \citep{Abdou2021-ox} and world knowledge \citep{Li2022-dt,Yun2023-po} to auditory perception \citep{Ngo2024-tw}. Interpreting these internal representations offers an alternative to traditional \ac{MIR} tagging approaches, helping to overcome limited dataset annotations and enabling potential inference-time control through representation editing. Recent work training probe classifiers on embeddings of music generative models shows that they capture both high-level musical concepts—such as genre, emotion, and instrument type \citep{Castellon2021-ry,Koo_undated-hp}—and lower-level music-theoretic features \citep{Wei2024-music}. Other studies explore how architecture and self-supervised learning influence such representations in music understanding tasks \citep{Won2023-sa,Li2023-dg}. Meanwhile, several works \citep{Zhao2025SteeringAM, Singh2025DiscoveringAS} have leveraged internal representations of musical concepts for activation steering \citep{turner2024steeringlanguagemodelsactivation}. Future research may further explore disentanglement in polysemantic features \citep{elhage2022toymodelssuperposition}, where individual neurons in a music generation models may be entangled with multiple musical features.

\subsection{Multimodality}\label{sec:multimodal}



There is a clear trend toward multimodal generative AI. While Section~\ref{sec:generative_models} discusses the rise of text-to-music systems, video-to-music generation remains relatively unexplored \citep{kang2024video2music, su2024v2meow}. Similarly, real-time, adaptive music generation presents promising opportunities for conditioned generation, though it faces challenges such as latency constraints and limited training datasets.

Joint embedding models, which represent multiple modalities in a shared semantic space, facilitate cross-modal retrieval, alignment scoring, and generation. These models are typically trained via contrastive learning, as in the vision-language model CLIP \citep{radford2021learning}, and analogous audio-text models like CLAP (Contrastive Language–Audio Pretraining) \citep{elizalde2023clap,wu2023large}, MuLan \citep{huang2022mulan}, and text-MIDI models such as CLaMP \citep{wu2023clamp}.

Despite these advances, multimodal representation learning for symbolic music remains underexplored. Recent work also shows that multimodal audio models can over-rely on language, potentially compromising audio fidelity \citep{weck2024muchomusic}. Future research may benefit from integrating symbolic, audio, video, and textual modalities, enabling cross-modal transformations and leveraging larger datasets to build more robust foundational representations.

\subsection{Efficiency}\label{sec:constraints}

Efficiency has become a central concern as music AI models grow in size and complexity. Smaller, faster models are not only more accessible to researchers with limited computational resources, but are also essential for latency-sensitive applications such as live music performance and interactive systems. Recent progress has been driven by techniques such as mixed-precision training, model quantization, and specialized hardware acceleration.

Lightweight models combining signal processing and deep learning have emerged as an active research area. PESTO \citep{riou2023pesto} achieves state-of-the-art fundamental frequency estimation with only 29k parameters by leveraging the transpositional invariance of the Constant-Q Transform (CQT), while Basic Pitch \citep{bittner2022lightweight} performs accurate pitch tracking with around 16k parameters using a harmonic CQT representation. RAVE \citep{caillon2021rave} enables real-time audio synthesis and style transfer, operating up to 80× faster than real time on a standard CPU through multi-band decomposition and Pseudo Quadrature Mirror Filters (PQMF) \citep{nguyen1994near}. Even in the text-to-song domain, relatively tiny models such as JAM \citep{liu2025jam} demonstrate that optimized architectures and high-quality data can efficiently achieve strong performance.

Many of the most efficient models achieve their performance by integrating of Digital Signal Processing (DSP) methods—such as Fourier variants and audio decomposition—with modern deep learning paradigms. Continued research at this intersection is likely to remain key for advancing efficient, real-time music AI systems. The next section discusses several applied areas that have seen rapid development with the rise of AI technologies.

\section{Selected music AI applications}\label{sec:applied}

We discuss a handful of selected applications that benefit from the recent developments in AI. We will briefly discuss a few of the latest systems as well as the challenges remaining and opportunities offered by these foundation models.

\subsection{Generative Models}\label{sec:generative_models}

While comprehensive surveys of generative music models exist \citep{herremans2017functional, ji2023survey, le2024natural, chen2025diffrhythm}, we briefly highlight recent systems and remaining challenges.

Text-conditioned audio-based generation has advanced rapidly \citep{melechovsky2023mustango, agostinelli2023musiclm, copet2024simple}, yet many still generate only short or instrumental-only fragments, though cascading approaches enable longer outputs \citep{schneider2023mo}. Open-source models are limited, with notable examples including MusicGen \citep{copet2024simple}, Stable Audio Open \citep{evans2024stableaudioopen}, Magenta Real-Time \citep{gdmlyria2025live}, and Mustango \citep{melechovsky2023mustango}; even fewer are trained solely on open data. Many commercial models, such as those from Suno and Udio, are unavailable for research and may rely on copyrighted datasets. 

Real-time music generation \citep{gdmlyria2025live} in particular with human interaction \citep{blanchard2025jam_bot, scarlatos2025realjamrealtimehumanaimusic} represents an exciting avenue for artist-in-the-loop musical performance. However, quality real-time generation remains constrained by hardware requirements. High-quality systems typically require GPUs—for instance, Mustango takes ~40 seconds to generate 10s of audio on an NVIDIA A100. Research into GPU-optimized architectures and efficiency improvements, such as the flow-based model JAM \citep{liu2025jam}, is crucial, especially with the rise of inference alignment techniques \citep{roy2025text2midi}. Furthermore, developing models that can run on-device in real-time, responsive to human interactivity, would greatly democratize music co-creation.

For practical adoption, generative tools must integrate with musicians’ familiar \acp{DAW}, typically via \ac{VST} plugins \citep{marrington2017composing}. However, this integration remains challenging because \acp{VST} are coded in C++ while most machine learning models use Python. Limited attempts have been made to bridge the gap between Python and \acp{VST} \citep{braun2021dawdreamer}. Such interfaces would enable these systems to become true tools for co-creation.

Singing voice synthesis presents another significant challenge, primarily due to its demand for large amounts of copyright-free training data. Recent models like JAM \citep{liu2025jam}, DiffRhythm$+$ \citep{chen2025diffrhythm}, SongBloom \citep{yang2025songbloom}, YuE \citep{yuan2025yue}, and Ace Step \citep{gong2025ace} demonstrate progress, but still require massive datasets (e.g., DiffRhythm$+$ trained on over one million songs), which are rarely openly available.

Finally, MIDI remains a popular format among professional music producers in DAW software. Current work for text-to-MIDI generation include MuseCoco \citep{lu2023musecoco} and Text2midi \citep{bhandari2025text2midi}, while \citep{donahue2024hookpad} leverages the HookTheory platform for human-AI co-creation in symbolic music. Many more symbolic music datasets have emerged, such as MidiCaps \citep{melechovsky2024midicaps} and Aria \citep{bradshaw2025aria} (transcribed audio recordings). However, high-quality data remains scarce, limiting the capabilities of generative MIDI models as discussed in Section~\ref{sec:datasets}. 
In the following subsections, we dive deeper into the various aspects of music generation systems.


\subsubsection{Evaluation Metrics}\label{sec:eval}
%

Evaluating music and audio systems remains challenging due to the subjective nature of human perception and the diversity of listener preferences. While numerous metrics exist for both audio (e.g., KL divergence, Fréchet Audio Distance  \citep{kilgour2018fr}) and symbolic systems \citep{dong2020muspy}, the lack of standardization across studies makes benchmarking and comparison difficult. Mauve Audio Divergence \citep{huang2025aligningtexttomusicevaluationhuman} seeks to improve previous systems like FAD by better aligning evaluation metrics with human preferences through pretrained foundation model representations.

To assess semantic alignment between music and text prompts, metrics like CLAP-score \citep{elizalde2023clap} are commonly used, though CLAP was not specifically trained for music-theory and finer-grained concepts. Alternative approaches have relied on feature predictors to compare generated music with input prompts \citep{melechovsky2023mustango}. In the symbolic domain, metrics such as musical interval statistics \citep{MakrisD2021ijcnn} and structure-based measures like COSIATEC compression ratio \citep{meredith2013cosiatec} are employed \citep{herremans2017morpheus}. For a comprehensive overview of objective evaluation metrics, readers are referred to \cite{ji2020comprehensive, kader2025survey}. 

Despite these efforts, objective metrics alone are often insufficient \citep{vinay2022evaluating}, and a growing consensus suggests that deeper musical insight is needed \citep{modrzejewski2023text}. As a result, subjective evaluation through listener surveys remains essential, though these require careful design to ensure meaningful results. Standardized listening tests are needed, and practices like cherry-picking examples should be avoided. The recent Music Arena platform \citep{kim2025musicarena} allows for users to generate and rate music across different models. \cite{AgresKR2016cie} offer a comprehensive survey of subjective evaluation techniques for creative music systems.

Recent work has shifted from low-level signal or embedding metrics toward measures that better capture perceived musical quality. SongEval \citep{yao2025songeval} introduces a dataset of over 2,000 full-length songs annotated by professional musicians across dimensions such as coherence, phrasing, and overall musicality. Similarly, Meta’s Audiobox-Aesthetics model \citep{tjandra2025aes} predicts perceived audio quality across speech, music, and environmental sounds, assessing attributes like production quality, complexity, and enjoyment. Together, these models move evaluation toward human-centered aesthetic judgment rather than purely objective or distributional measures. Their quantitative nature also enables integration into Direct Preference Optimization (DPO) \citep{dong2024its} or inference-time output selection frameworks \citep{roy2025text2midi}, where aesthetic scores can guide ranking or fine-tuning toward more musically appealing results.

\subsubsection{Controllability}\label{sec:control}

Controllable generation aims to guide AI models toward outputs with user-specified attributes and has seen progress in music AI. Traditional MIDI-based models allow conditioning on musical features or emotions \citep{MakrisD2021ijcnn}, while recent text-based models primarily provide high-level control (e.g., genre, mood). Only a few models, such as Mustango, enable incorporation of music-theoretic constraints, allowing generation with specific chord sequences, tempos, or keys. 

Text control is typically global, but fine-grained, time-varying control over low-level aspects like dynamics or tempo requires more intuitive interfaces. Music ControlNet \citep{Wu2023-ri} addresses this by combining global controls (mood, genre) with time-varying controls (melody, rhythm, dynamics), while other works adopt symbolic, compositional, or rule-based methods \citep{Huang2024-ek,Thickstun2023-vm}. 

Emotion is another key dimension of control. MIDI-based models have explored emotion conditioning \citep{herremans2017morpheus, MakrisD2021ijcnn}. While LLM-based systems often account for emotion implicitly, explicit emotion control remains underexplored, yet it holds potential for applications in mental health and personalized music generation \citep{agres2021music}.

Iterative music editing remains a promising direction for interactive, controllable music co-creation. Models such as MusicMagus \citep{zhang2024musicmagus} and Instruct-MusicGen \citep{zhang2024instruct} are equipped with large multimodal models that allow the user to change the instrument, genre, mood, or add an additional stem on top of the existing accompaniment. To train such models to generate and edit musical stems, such as MusicGen-Stem \citep{rouard2025musicgen}, researchers often build upon datasets for source separation, which can be augmented for stem generation.

Looking forward, substantial room for improvement remains in enabling musicians and composers to truly co-create with AI through more descriptive, abstract, temporally aware, and theory-informed instructions, e.g. `music for a gym workout, starting with an ascending melody, introducing a key change up two semitones, and ending on a C-note.'.

\subsection{Music production and mixing}

Recent advances in AI-based music production have introduced new approaches to \ac{AMM} \citep{SteinmetzC2021icassp, MartinezRamirezMA2022ismir}. However, progress is limited by the lack of publicly available training data—namely, paired dry (anechoic) recordings and their mastered counterparts. Expert annotations from professional audio engineers could further improve supervised learning methods. Moreover, several studies note the shared objectives between music source separation and \ac{AMM}, suggesting that joint optimization could yield more tractable and interpretable mixing and separation strategies \citep{YangH2022icassp}.

Recently, text-controlled music production models have emerged. Notably, \cite{chu2025text2fx}’s text2fx introduced text-guided audio effects processing, while SonicMaster \citep{melechovsky2025sonicmaster}, a flow-based model, represents the first all-in-one text-driven system for music restoration and mastering. To overcome the lack of training data, SonicMaster artificially degrades production-quality recordings using 16 types of distortions, enabling it to respond to natural-language commands such as “bring out the bass” or “reduce the reverb.”

Beyond mixing and mastering, rendering human-like expressiveness from MIDI remains a key challenge. This spans expressive performance generation from quantized MIDI \citep{cancino2018computational} and the design of neural sound fonts, from models emulating analog gear such as tube amplifiers \citep{damskagg2019deep} to those directly synthesizing instrument sounds from symbolic input \citep{castellon2020towards}. Although research in this area has slowed, there remains a clear need for end-to-end systems that transform MIDI into expressive, fully mastered audio by unifying synthesis and mastering.

\subsection{Music Captioning}

Music captioning--the task of generating natural-language descriptions of musical audio--is also emerging as a challenge in music AI. Early work, such as MusCaps \citep{manco2021muscaps} introduced encoder–decoder architectures that produce a descriptive sentence (e.g., instrumentation, mood, genre) rather than classical tags. Recently, MusiLingo \citep{deng2023musilingo} bridges pretrained music-audio encoders (e.g., MERT) with large language models (LLMs) via a projection layer, enabling music caption generation and query-response interactions (“what is the mood?”, “which instruments?”, etc.). SonicVerse \citep{chopra2025sonicverse} advances the field by adopting a multi‐task architecture: besides caption generation, it jointly predicts auxiliary musical feature labels (key, vocal presence, instrumentation, genre) and injects these feature token projections into a language model to produce richer captions. Meanwhile, large multimodal audio‐language models such as Qwen3-Omni \citep{xu2025qwen3} suggest the potential for generalized audio captioning systems that can handle music, speech and ambient sound in a unified way, though their specialization for music captioning remains underexplored. 

Despite this progress, considerable challenges remain. Datasets are still small relative to those in vision or language domains, and standard captioning metrics (BLEU, ROUGE, METEOR) correlate poorly with human judgments in the music domain \citep{lee2024captioning}. Furthermore, musical audio presents overlapping layers (multiple instruments, effects, production artifacts, and expressive timing) and a time element that complicate capturing high-level semantics in text. Overcoming these bottlenecks and scaling captioning systems across large song catalogs could enable richly annotated natural‐language descriptions that improve both retrieval and the training of generation systems.

\subsection{Music transcription}

Music transcription—the automatic conversion of audio recordings into symbolic representations such as MIDI—is often described as a “holy grail” of the \ac{MIR} field. While considerable progress has been made, particularly for monophonic recordings or isolated instruments (e.g., note-level transcription of solo piano), accurate transcription of polyphonic, multi-instrument performances remains a major challenge. For instance, \cite{wang2018polyphonic} proposed a note-based music language model combined with an acoustic front-end for polyphonic piano transcription, while \cite{simon2022scaling} demonstrated that scaling training data via mixtures of monophonic recordings improved multi‐instrument polyphonic transcription. Overcoming this barrier would enable transcription of millions of existing audio songs into accurate, richly annotated polyphonic MIDI (or equivalent symbolic) files. Such a dataset would be a transformative resource, enabling the training of large-scale generative MIDI models such as Text2midi \citep{bhandari2025text2midi}.

\subsection{Source separation}

In recent years, neural networks have achieved state-of-the-art performance in music source separation \citep{StollerD2018waveunet, DefossezA2019demucs}.
Models such as Spleeter \citep{HennequinR2020spleeter} have been open-sourced, while proprietary tools like LALAL.AI have seen widespread commercial adoption.
Recent research has explored novel model architectures such as band-split RNN \citep{LuoY2023ieeeacmtaslp}, transformers \citep{RouardS2023icassp}, as well as new training paradigms (e.g., augmenting training data \citep{PonsJ2024icassp} or modifying the loss function \citep{
SawataR2023whole}).
Despite these advances, separating certain instruments, such as distinct singing voices or synthesizers, remains challenging.
Accurate source separation would not only empower music producers but also enable the generation of large-scale stem datasets for training music generation models.

\subsection{Music Discovery}\label{sec:discovery}

The integration of recommender systems, deep learning, and music information retrieval has transformed how users discover music, enabling highly personalized suggestions based on listening habits and preferences. Yet this personalization introduces challenges, such as the echo chamber effect, where algorithms reinforce existing tastes and limit exposure to new genres, artists, or cultures. Long-term user engagement remains difficult to capture \citep{wang2022surrogate}, while recent surveys emphasize the growing importance of fairness and transparency in recommender design \citep{wang2023survey}. \cite{zangerle2022evaluating} highlight the need for long-tail recommendations that promote older or lesser-known tracks, and \cite{knees2020intelligent} analyze discovery interface principles that balance metadata quality with contextual and artist-centric cues.

Recent advances extend beyond conventional collaborative filtering toward multimodal and conversational recommendation. Models such as CLAP \citep{elizalde2023clap} and MuLan \citep{huang2022mulan} align text and audio representations, allowing users to search and discover music through natural-language descriptions rather than fixed tags. Emerging LLM-driven recommenders (e.g., ChatMusicRec \citep{kim2024large}, LLM-Rec \citep{liu2023llmrec}) act as dialogue agents that elicit user intent and justify their suggestions, providing more transparent and exploratory discovery experiences.

\subsection{Music Performance}\label{sec:performance}

Using neural networks as live musical instruments introduces a new paradigm in musical creativity, where artificial intelligence collaborates with human musicians to generate novel sounds and compositions.
For instance, RAVE \citep{caillon2021rave} has been used to perform drum rhythms through integration with Max\footnote{\url{https://cycling74.com/products/max}} and motion sensor controls, allowing to create rhythms through expressive gestural movements in the air. Recently, researchers have developed an AI symbolic music jam bot, based off of Anticipatory Music Transformer \citep{Thickstun2023-vm}, to perform a live concert in real-time with GRAMMY-winning artist Jordan Rudess \citep{blanchard2025jam_bot}.
The sampling and looping capabilities of VampNet \citep{garcia2023vampnet} have been used for creating live experimental soundscapes with a live instrumentalist through dynamic interplay between the musician with the model.

AI models can also adapt in real-time to human performer inputs, enabling otherwise physically impossible instruments.
The HITar exemplifies this approach: an augmented acoustic guitar played with hybrid-percussive technique, morphed and extended in real-time by AI-generated tabla sounds \citep{martelloni2023real}.
We consider the usage of AI to create entirely new, previously undiscovered means of musical expression an especially exciting and worthwhile endeavor.

\subsection{Music Education}\label{sec:education}

Although applications of AI in music education remain under-explored, most approaches aim either to assist students directly in musical practice or to enhance the accessibility and therapeutic aspects of learning music. These include systems that provide feedback, generate personalized learning materials, or offer new modes of musical interaction. \cite{gover2022music} propose difficulty-level conversion, generating easier or harder versions of piano arrangements, while pitch and beat tracking models can support vocal or instrumental practice. \cite{morsi2023sounds} detect conspicuous mistakes in piano performances without requiring a score for comparison, and \cite{balliauw2017variable} automate the generation of piano fingerings for students. 

More recently, large language models and multimodal AI systems have been explored as intelligent tutors and performance coaches.
\cite{pond2025teaching} show that LLMs can reason about music theory through in-context and chain-of-thought prompting, while \cite{zhang2025llaqo} propose LLaQo, a multimodal coach providing expressive feedback on performance.  Together, these studies suggest that adaptive, conversational, and multimodal AI systems could soon play a central role in personalized and creative music learning.

\section{Responsible Music AI}\label{sec:responsible}

\subsection{Datasets}\label{sec:datasets}

The diversity of MIR tasks creates significant opportunities for developing new datasets to advance the field. Historically, audio datasets have varied in format: some provide small collections of raw audio, while others distribute only extracted features or track IDs \citep{won2021music}. MIDI datasets represent another important modality, enabling research on note sequences, timing, and musical structure, but often require different processing approaches than audio. While it is impossible to discuss all MIR datasets comprehensively, we outline several key challenges below. 

Musical stems are a particularly sought-after data source, as music producers naturally work with stems in their DAW, and such datasets can facilitate the training of stem-generation and source separation models. MIDI datasets can be synthesized per track, enabling stem-separated audio. However, rendering MIDI introduces significant artifacts and limited expressivity \citep{manilow2019cutting}. 
Real instrument datasets like MUSDB18 \citep{musdb18}, though popular, are relatively small (approximately 10 hours) and now serve more as benchmarks than as comprehensive training resources.

To address coverage gaps in instruments and genres, researchers increasingly turn to larger open-domain datasets such as FMA, which offers over 106,000 tracks or roughly \SI{343}{days} of Creative Commons-licensed audio. However, these often lack individual instrument stems, requiring self-supervised or unsupervised methods for MSS. 

A major limitation persists in legally obtainable training data for text-to-music generation. Captioned music datasets are scarce, with only a few copyright-cleared options available—MusicCaps \citep{agostinelli2023musiclm}, MusicCaps-LM \citep{doh2023lp}, MusicBench \citep{melechovsky2023mustango}, and JamendoMaxCaps \citep{roy2025jamendomaxcaps}. 
For MIDI, the recently released MidiCaps \citep{melechovsky2024midicaps} is the only dataset offering full-length files with detailed text descriptions. This scarcity has limited progress in text-to-MIDI systems, though recent models like MuseCoco \citep{lu2023musecoco} attempt to overcome these gaps by extracting features from input captions. In all these datasets, it is important to also consider cultural (particularly Western) bias as well as the annotation quality (e.g. LLM-annotated captions versus human-annotated). 

In summary, dataset limitations remain a significant bottleneck. There is a substantial need for copyright-cleared stem and MIDI datasets, annotated with text captions including music-theory descriptions (to enable reasoning) and general descriptions such as instrumentation, emotion, and mood.


\subsection{Respecting artist rights} \label{sec:copyright}


Over the years, disruptive technologies—from cable television to MP3 players and streaming services—have repeatedly clashed with copyright holders, often resulting in legal battles that ultimately favored technological progress. Today, generative AI models are advancing at a pace that far exceeds the evolution of laws and policies \citep{SamuelsonP2023science}. 
Recent litigation has focused on large language models reproducing copyrighted content, such as song lyrics, from both licensed and web-scraped training data. For instance, Anthropic was sued by major music publishers for alleged infringement, while organizations like ASCAP are advocating for legislation that prioritizes human creators, consent, compensation, credit, transparency, and global consistency \citep{Poritz_2023}.

Copyright remains a central concern, as using licensed music tracks for model training--especially for commercial purposes—poses significant legal and ethical challenges. While copyright protections support artists and labels, they can also hinder the development of new music technologies. There is a growing need for equitable, easy-to-apply attribution mechanisms for music data and metadata \citep{herremans2025royalties}. Research into frameworks for attribution, fairness, and openness evaluation is increasingly important to balance the interests of creators and technologists. Potential solutions include robust melody and timbre similarity detection algorithms, which could facilitate legal data usage and profit-sharing without stifling research \cite{herremans2025royalties, lu2025melodysim}. 

In addition, deep learning-based watermarking \citep{roman2024proactive, chen2023wavmark}, originally designed for speech, is now being adapted for music. Data poisoning methods such as Glaze \citep{shan2023glaze} and Nightshade \citep{shan2024nightshade} have also shown promising results in the image domain, by applying subtle modifications to images that render the data ineffective for training specific generative models. 
Early research has begun exploring similar techniques for music \citep{barnett2024audiodata}. The development of new technologies for protecting artist rights is emerging as an important research topic.

\section{Conclusion}\label{sec:conclusion}

\textbf{\textit{So what should I work on now?}}
Although the history of AI and music extends as far back as 1950s, algorithms developed in just the last few years have achieved unparalleled capabilities, amplifying the now mainstream dialogue about AI.
The technologies and software discussed in this survey have already begun to impact music educators, producers, performers, record labels, and artists at every level.

Our discussion reveals that this rapid progress has introduced new challenges and opportunities unique to the field of \ac{MIR}.
From ethical datasets and copyright attribution, to multimodal controllable systems, real-time generative modeling, and novel methods for music production, remixing, and editing, there has never been a better time to become a music AI researcher.

\section{Acknowledgments}
This work has received support from SUTD's Kickstart Initiative under grant number SKI 2021\_04\_06 and MOE T2.
We acknowledge the use of ChatGPT for grammar improvements.




\bibliography{template/references}  
\end{document}